\overfullrule=0pt
\input harvmac


\def\a{{\alpha}}

\def\l{{\lambda}}
\def\lb{{\overline\lambda}}
\def\b{{\beta}}

\def\g{{\gamma}}

\def\d{{\delta}}
\def\e{{\epsilon}}

\def\half{{1\over 2}}
\def\p{{\partial}}

\def\t{{\theta}}

\def\bar{\overline}
\def\({\left(}
\def\){\right)}

\def\cF{ {\cal F} }


\Title{\vbox{\hbox{IFT-P.024/2006 }}}
{\vbox{
	\centerline{\bf Some Superstring Amplitude Computations  }
	\centerline{\bf with the Non-Minimal Pure Spinor Formalism }
 }
}

\bigskip\centerline{Nathan Berkovits\foot{email: nberkovi@ift.unesp.br}
                  and Carlos R. Mafra\foot{email: crmafra@ift.unesp.br}}

\bigskip
\centerline{\it Instituto de F\'\i sica Te\'orica, State University of
S\~ao Paulo}
\centerline{\it Rua Pamplona 145, 01405-900, S\~ao Paulo, SP, Brasil}

\vskip .3in 
We use the non-minimal pure spinor formalism to compute in a 
super-Poincar\'e covariant manner the four-point
massless one and two-loop open superstring amplitudes, and  
the gauge anomaly of the six-point one-loop amplitude.  
All of these amplitudes are expressed as integrals of ten-dimensional
superfields
in a ``pure spinor superspace'' which involves five $\t$ coordinates
covariantly contracted with three pure spinors. The bosonic
contribution to these amplitudes agrees with the standard results, and
we demonstrate identities which show how
the $t_8$ and $\e_{10}$ tensors naturally
emerge from integrals over pure spinor superspace.

\vskip .3in

\Date {July 2006}


\newsec{Introduction}

Although much has been learned about superstring amplitudes
using the Ramond-Neveu-Schwarz (RNS) formalism, the need to sum over spin
structures obscures the role of spacetime supersymmetry. Using the light-cone
Green-Schwarz (GS) formalism, one can easily compute four-point tree and
one-loop amplitudes with half of the supersymmetry manifest. But higher-point
and higher-loop
amplitudes are more difficult to compute in this light-cone formalism,
especially amplitudes that involve
the ten-dimensional $\e$ tensor. Although a covariant version of the
GS formalism has recently been developed by Lee and Siegel
\ref\LeePAtwo{
        K.~Lee and W.~Siegel,
        ``Conquest of the ghost pyramid of the superstring,''
        arXiv:hep-th/0506198.
}
\ref\LeePA{
	K.~Lee and W.~Siegel,
        ``Simpler superstring scattering,''
        arXiv:hep-th/0603218.
}, this covariant
GS formalism has not been used to compute higher-loop amplitudes or
amplitudes involving the $\e$ tensor.

Over the last six years, a manifestly super-Poincar\'e covariant
superstring formalism
has been developed which involves bosonic ghost variables $\l^\a$
satisfying the pure spinor constraint $\l\g^m\l=0$
\ref\superpoincare{
	N.~Berkovits,
        ``Super-Poincare covariant quantization of the superstring,''
        JHEP {\bf 0004}, 018 (2000)
        [arXiv:hep-th/0001035].
}.
Tree amplitudes and one-loop four-point amplitudes were computed 
in \ref\PSmulti{
	N.~Berkovits,
        ``Multiloop amplitudes and vanishing theorems using the pure spinor
        formalism for the superstring,''
        JHEP {\bf 0409}, 047 (2004)
        [arXiv:hep-th/0406055].
} using
a ``minimal'' version of the formalism,  
and these computations
were later extended to two-loop four-point amplitudes in 
\ref\BerkovitsDF{
	N.~Berkovits,
        ``Super-Poincare covariant two-loop superstring amplitudes,''
        JHEP {\bf 0601}, 005 (2006)
        [arXiv:hep-th/0503197].
}
and to $d=11$
one-loop computations in 
\ref\AnguelovaPG{
	L.~Anguelova, P.~A.~Grassi and P.~Vanhove,
        ``Covariant one-loop amplitudes in D = 11,''
        Nucl.\ Phys.\ B {\bf 702}, 269 (2004)
        [arXiv:hep-th/0408171].
}. 
When all external states are bosons, these amplitudes were
shown in
\ref\vallilo{B.~C.~Vallilo and N.~Berkovits, ``Consistency of 
super-Poincare covariant superstring tree amplitudes,''
        JHEP {\bf 0007}, 015 (2000) [arXiv:hep-th/0004171].
}
\ref\PSequivI{
	C.~R.~Mafra,
        ``Four-point one-loop amplitude computation in the pure spinor formalism,''
        JHEP {\bf 0601}, 075 (2006) [arXiv:hep-th/0512052].
}	
\ref\PSequivII{
	N.~Berkovits and C.~R.~Mafra,
        ``Equivalence of two-loop superstring amplitudes in the pure spinor and RNS
        formalisms,''
        Phys.\ Rev.\ Lett.\  {\bf 96}, 011602 (2006)
        [arXiv:hep-th/0509234].  
}
to coincide with the standard RNS result.

All of these amplitudes are expressed as integrals of superfields
in ``pure spinor superspace'' which, in $d=10$, 
involves five fermionic $\t$ coordinates
covariantly contracted with three bosonic pure spinors. 
When all superfields are on-shell, the superspace integrands are
annihilated by the pure spinor BRST operator $Q=\l^\a D_\a$. As shown in
\superpoincare, this implies that the amplitude expressions are
invariant under all sixteen $d=10$ supersymmetries even if the
pure spinor superspace only involves five $\t$'s.

More recently, a non-minimal version of the pure spinor formalism has
been developed which involves both a pure spinor $\l^\a$ and its complex
conjugate $\bar\l_\a$ \ref\NMPS{
	N.~Berkovits,
        ``Pure spinor formalism as an N = 2 topological string,''
        JHEP {\bf 0510}, 089 (2005)
        [arXiv:hep-th/0509120].
}. The amplitude prescription using the non-minimal
version is considerably
simpler than in the minimal version since there are no picture-changing
operators and Lorentz invariance is manifest at all stages in the computation.
Furthermore, the amplitude prescription in the non-minimal formalism
can be related to the
prescription in topological string theory where the $b$ ghost
is replaced by a composite operator.

For tree amplitudes, it is trivial to show that the minimal and non-minimal
pure spinor formalisms give the same answers. But for loop amplitudes,
there are some differences between the minimal and non-minimal computations
which makes it non-trivial to prove their equivalence.
In the first part of this paper, 
the non-minimal pure spinor formalism will be
used to re-compute the massless four-point one-loop and two-loop amplitudes
and equivalence with the minimal computations will be proven.
In terms of integrals over
pure spinor superspace, the kinematic factors in these one-loop and
two-loop amplitudes will be shown to be proportional to 
\eqn\kine{ K_{1-loop} = \langle (\l A)(\l\g^m W)(\l\g^n W) \cF_{mn} \rangle,}
$$K_{2-loop} = \langle (\l \g^{mnpqr}\l)\cF_{mn}\cF_{pq}\cF_{rs}(\l\g^s W) 
\rangle,$$
where 
$A_\a$, $W^\a$, and $\cF_{mn}$ are the spinor gauge superfield, 
spinor superfield-strength, and vector superfield-strength of
super-Yang-Mills, and 
the pure spinor
measure factor $\langle ~\rangle$ is defined
such that 
$\langle (\l\g^m\t)(\l\g^n\t)(\l\g^p \t)(\t\g_{mnp}\t) \rangle=1$.
Using the super-Yang-Mills equations of motion, it is easy to check
that the integrands in \kine\ are annihilated by $\l^\a D_\a$, so
these kinematic factors are supersymmetric. 

The non-minimal formalism will then be used to compute in a 
supersymmetric manner the
gauge variation of the massless six-point one-loop amplitude
in Type-I superstring theory. Since this computation involves the
ten-dimensional $\e$ tensor, it has never been performed using the
light-cone GS formalism. After expressing the gauge variation of the
six-point amplitude as a term at the boundary of moduli space, it
will be shown that the anomaly 
is proportional to the pure spinor superspace integral
\eqn\anomw{
K_{anomaly} = \langle (\l\g^m W)(\l\g^n W) (\l\g^p W)(W\g_{mnp} W) \rangle,}
whose purely bosonic contribution is the standard $\e_{10} F^5$ term.
 
Further investigation upon the appearance of $\e_{10}$ in \anomw\
led us to the discovery of
a pure spinor superspace integral, namely,
$$
\langle (\l\g^r W^1)(\l\g^s W^2) (\l\g^t W^3)(\t\g^m\g^n\g_{rst} W^4) \rangle,
$$
from which the $t_8$ and $\e_{10}$ tensors naturally emerge in a unified manner,
in the form $\eta^{mn}t_8^{m_1n_1{\ldots}m_4n_4}-{1\over 2}\e_{10}^{mnm_1n_1{\ldots}
m_4n_4}$.
This differs from the RNS formalism where the $t_8$ and $\e_{10}$
tensors come from different spin structures. It may be possible
that this pure spinor superspace integral is related to the five-point one-loop
amplitudes involving the 
heterotic $\e_{10}B {\rm tr}F^4$ and Type IIA $\e_{10}t_8 B R^4$ terms, 
which would be useful
for finding the supersymmetric completions of these terms.

It is interesting to compare these computations using the pure spinor
formalism with the recently developed method of Lee and Siegel for
computing one-loop amplitudes.
The method of Lee and Siegel is based on the ``ghost pyramid'' covariant
quantization of the Green-Schwarz superstring, in which the BRST
operator has a complicated structure involving an infinite set of ghosts
\LeePAtwo.
However, the vertex operators in the Lee-Siegel formalism are relatively
simple and have a very similar structure to the integrated vertex operator
in the pure spinor formalism. 

In the one-loop computations performed in \LeePA\
using the Lee-Siegel
method, all vertex operators are integrated and there are no picture-changing
operators. Furthermore, there is no superspace integration using this 
method so the amplitudes are expressed in terms of the component fields.
This is the analog of the ${\cal F}_1$ picture in the RNS formalism
where all vertex operators are in the zero picture.

On the other hand, in the
one-loop computations using the pure spinor formalism,
one of the vertex operators is unintegrated, the $b$ ghost
is a composite operator playing the role of a picture-raising operator,
and the amplitudes are expressed as integrals over pure spinor superspace.
This is the analog of the ${\cal F}_2$ picture in the RNS formalism
where the unintegrated vertex operator is in the $-1$ picture and
the picture-raising operator is inserted on top of the $b$ ghost.

For certain one-loop computations such as the four-point and
five-point massless amplitudes
computed in \LeePA, there is no disadvantage in
treating all vertex operators in integrated form. 
However, for the anomaly computation
presented here, it is definitely more convenient to leave one unintegrated
vertex operator in a ``different picture'' from
the integrated vertex operators.
It would be interesting to see how
to compute this anomaly using the Lee-Siegel method, and if one needs to
introduce some analog of picture-changing operators.

In section 2, we review the non-minimal amplitude prescription
for one-loop and two-loop amplitudes. In section 3, we compute
the massless four-point one-loop and two-loop amplitudes and
show agreement with the computations using the minimal formalism.
In section 4, we compute the gauge variation of the massless six-point
one-loop amplitude. In section 5, we explain how $t_8$ and $\e_{10}$
tensors naturally emerge from the integration over pure spinor superspace.
And in appendix A, we list all the pure spinor superspace identities
used in this paper and present two other representations for
$t_8$ and $\e_{10}$ tensors using pure spinors.

\newsec{Non-Minimal Amplitude Prescription}

The prescription in the non-minimal pure spinor formalism
for computing $N$-point one-loop
and two-loop 
scattering amplitudes is given by \NMPS
\eqn\onep{
{\cal A}_{1-loop} = \int d\tau \langle {\cal N}~ (\int dw \mu(w) b(w))
V_1(z_1) \prod_{r=2}^{N}\int dz_rU_r(z_r) \rangle,
}
and
\eqn\twop{
{\cal A}_{2-loop} = \int d\tau_1d\tau_2d\tau_3 \langle
{{\cal N} ~
\prod_{s=1}^3(\int dw_s \mu_s(w_s) b(w_s))
             \prod_{r=1}^{N}\int dz_rU_r(z_r)}\rangle.}
where $\tau_i$ are the Teichmuller parameters, $\mu_i$
are the Beltrami differentials, $V_r$ and $U_r$ are the
unintegrated and integrated vertex operators,
and $\langle~~\rangle$ denotes
the functional integral over the Green-Schwarz-Siegel
fields $[x^m, \t^\a, d_\a]$, over the pure spinor ghosts $\l^\a$
and their conjugate momenta $w_\a$,
and over the non-minimal fields $[\lb_\a, r_\a]$ and
their conjugate momenta $[\bar w^\a, s^\a]$.

As in topological string theory, the
$b$-ghost is a composite operator satisfying $\{Q,b\}=T$ where 
$T$ is the stress-tensor,
and has the explicit form
\eqn\bghost{
b = s^{\a}\p\lb_{\a} + {2\Pi^m(\lb\g_m d)-N_{mn}(\lb\g^{mn}\p\t)-J(\lb\p\t)
-(\lb\p^2\t) \over 4(\lb\l)}
}
$$
+{ (\lb\g^{mnp}r)(d\g_{mnp}d+24N_{mn}\Pi_p) \over 192(\lb\l)^2}
-{(r\g^{mnp}r)(\lb\g_m d)N_{np} \over 16(\lb\l)^3}
+{(r\g^{mnp}r)(\lb\g_{pqr}r)N_{mn}N^{qr} \over 128(\lb\l)^4}
$$
where $\Pi^m = \p x^m + \half (\t \g^m \p\t)$ is the supersymmetric momentum
and $N_{mn} = \half (w\g_{mn}\l)$ and $J = \l w$ are the pure spinor
Lorentz and ghost currents. 

Integration over the zero modes of the bosonic and fermionic
worldsheet fields naively gives
$0/0$, so it is necessary to insert a BRST-invariant operator 
${{\cal N}= e^{\{Q, \chi\}}}$
which regularizes this 
zero mode integration. Since ${\cal N}=1 + \{Q,\Omega\}$, the choice
of $\chi$ does not affect the scattering amplitude. A convenient choice is
$\chi = -\lb_\a \t^\a - \sum_{I=1}^g (\half N_{mn}^I (s^I\g^{mn}\lb)
+ J^I (s^I\lb)  ) $,
which implies that
\eqn\reg{
	{\cal N}= \exp ( -\lb_\a\l^\a -r_\a \t^\a)
}
$$
\exp(~
\sum_{I=1}^g [~-\half N_{mn}^I \bar N^{mn I} - J^I \bar J^I ~~
- {1\over 4} (s^I \g_{mn}\lb)(\l\g^{mn} d^I) + (s^I \lb)(\l d^I)]~), $$
where $[N_{mn}^I, J^I, \bar N_{mn}, \bar J^I, d_\a^I, s^{ I \a }]$ 
denote the $g$ zero modes of these
spin one fields on a genus $g$ surface. 

Finally, for massless external states, the unintegrated vertex operator
is $V = \l^\a A_\a$ and the integrated vertex operator is 
$$ 
U = \p\t^{\a} A_{\a} +\Pi^m A_m + d_{\a}W^{\a} 
+{1\over 2}N^{mn} \cF_{mn}.
$$
The $ [A_{\a}, A_n, W^\a,  \cF_{mn}]$
superfields describe super-Yang-Mills theory \ref\SYM{
	E.~Witten,
        ``Twistor - Like Transform In Ten-Dimensions,''
        Nucl.\ Phys.\ B {\bf 266}, 245 (1986).
\semi
	W.~Siegel,
        ``Superfields In Higher Dimensional Space-Time,''
        Phys.\ Lett.\ B {\bf 80}, 220 (1979).
} and have
the $\t$-expansions \ref\thetaSYM{
	H.~Ooguri, J.~Rahmfeld, H.~Robins and J.~Tannenhauser,
        ``Holography in superspace,''
        JHEP {\bf 0007}, 045 (2000)
        [arXiv:hep-th/0007104].
\semi
	P.~A.~Grassi and L.~Tamassia,
        ``Vertex operators for closed superstrings,''
        JHEP {\bf 0407}, 071 (2004)
        [arXiv:hep-th/0405072].
\semi
	G.~Policastro and D.~Tsimpis,
        ``R**4, purified,''
        arXiv:hep-th/0603165.
}
$$
A_{\a}(x,\t)={1\over 2}a_m(\g^m\t)_\a -{1\over 3}(\xi\g_m\t)(\g^m\t)_\a
-{1\over 32}F_{mn}(\g_p\t)_\a (\t\g^{mnp}\t) + \ldots
$$
$$
A_{m}(x,\t) = a_m - (\xi\g_m\t) - {1\over 8}(\t\g_m\g^{pq}\t)F_{pq}
         + {1\over 12}(\t\g_m\g^{pq}\t)(\p_p\xi\g_q\t) + \ldots
$$
$$
W^{\a}(x,\t) = \xi^{\a} - {1\over 4}(\g^{mn}\t)^{\a} F_{mn}
           + {1\over 4}(\g^{mn}\t)^{\a}(\p_m\xi\g_n\t)
	   + {1\over 48}(\g^{mn})^{\a}(\t\g_n\g^{pq}\t)\p_m F_{pq} 
	   + \ldots
$$
$$
\cF_{mn}(x,\t) = F_{mn} - 2(\p_{[m}\xi\g_{n]}\t) + {1\over
4}(\t\g_{[m}\g^{pq}\t)\p_{n]}F_{pq} + {\ldots},
$$
where $a_m(x)$ and $\xi^{\a}(x)$ describe the gluon and gluino 
fields, $F_{mn} = 2\p_{[m} a_{n]}$, and $\ldots$ involve derivatives
of $a_m$ and $\xi^\a$.

To compute the functional integral over the worldsheet fields, one first
uses the free field OPE's to integrate out the non-zero modes. Note that
as in topological string theory, computation of the partition function for 
the non-zero
modes is trivial because of cancellations between bosonic and
fermionic fields of equal spin.
The worldsheet zero
modes are then integrated out using the measure factors described in
\NMPS\ and the regulator ${\cal N}$ of \reg.

\newsec{Four-Point One-Loop and Two-Loop Computations}

As was shown in \PSmulti\ and \PSequivII\ 
using the minimal pure spinor formalism,
the kinematic factors for the massless four-point one-loop and two-loop
amplitudes are proportional to the pure spinor superspace integrals 
\eqn\kinet{ K_{1-loop} = \langle (\l A)(\l\g^m W)(\l\g^n W) \cF_{mn} \rangle
,}
\eqn\kineu{
K_{2-loop} = \langle (\l \g^{mnpqr}\l)\cF_{mn}\cF_{pq}\cF_{rs}(\l\g^s W) 
\rangle,}
where 
$A_\a$, $W^\a$, and $\cF_{mn}$ are the spinor gauge superfield, 
spinor superfield-strength, and vector superfield-strength of
the four external super-Yang-Mills multiplets, the expressions of \kinet\
and \kineu\ are summed over permutations of the four external
superfields, and
the pure spinor
measure factor $\langle ~\rangle$ is defined
such that 
$\langle (\l\g^m\t)(\l\g^n\t)(\l\g^p \t)(\t\g_{mnp}\t)\rangle =1$.
In \PSequivI\ and \PSequivII, the purely bosonic contributions to these 
pure spinor superspace integrals where shown to correctly reproduce the
$t_8$ index contractions of the four Yang-Mills field-strengths.

It will now be shown that the non-minimal computation of the four-point
massless one-loop and two-loop amplitudes contains the same kinematic
factors as in \PSequivI\PSequivII. Since the 
moduli space part
of the amplitude computations 
in the minimal and non-minimal formalisms is the same,
this proves the equivalence of the two prescriptions for these
amplitudes.

\subsec{One-loop computation}

Using the one-loop prescription of \onep, the regulator ${\cal N}$ of
\reg\ can
provide a maximum of eleven $d_\a$ zero modes, which are multiplied
by the eleven $s^\a$ zero modes. So the remaining five
$d_\a$ zero modes must come either from the vertex operators or from
the single $b$ ghost. Since the three integrated vertex operators
can provide at most three $d_\a$ zero modes through the terms
$(W^\a d_\a)$, the single $b$ ghost of \bghost\
must provide two $d_\a$ zero modes
through the term
\eqn\twod{{{(\lb\g^{mnp} r)(d\g_{mnp}d)}\over{192(\l\lb)^2}}.}

After integrating over the zero modes of the dimension one fields
$(w_\a, \bar w^\a, d_\a, s^\a)$ using the measure factors described
in \NMPS, one is left with an expression proportional
to 
\eqn\loopone{\int d^{16}\t \int [d\l][d\lb][dr] (\l\lb)^{-2} (\l)^4 
(\lb\g^{mnp} r) A W W W \exp (-\l\lb - r\t)}
\eqn\looptwo{=\int d^{16}\t
\int [d\l][d\lb][dr] \exp (-\l\lb - r\t)(\l\lb)^{-2} (\l)^4 
(\lb\g^{mnp} D) A W W W }
where
$D_\a= {\p\over{\p\t^\a}} + (\g^m\t)_\a \p_m$ is the usual
superspace derivative and the index contractions on
\eqn\indexc{(\l)^4 (\lb\g^{mnp} D) A W W W }
have not been worked out. 
Note that \looptwo\ is obtained from
\loopone\
by writing $ r_\a \exp(-r\t)= 
{\p\over{\p\t^\a}} \exp(-r\t)$, integrating by parts with respect
to $\t$, and using conservation of momentum
to ignore total derivatives with respect to $x$. 
Furthermore, the factor of $(\l)^4$ in \loopone\
comes from the $\l$ in the
unintegrated vertex operator,
the $11$ factors of $\l$ and $\lb$ which multiply the zero modes of
$d_\a$ and $s_\a$ in ${\cal N}$, the factor of $(\l)^{-8}(\lb)^{-8}$
in the measure factor of $w_\a$ and $\bar w^{\a}$, and the factor
of $(\lb)^{-3}$ in the measure factor of $s^\a$.

Fortunately, it is easy to show there is a unique Lorentz-invariant
way to contract the indices in \indexc. To show this, first choose
a Lorentz frame in which the only non-zero component of $\l^\a$
is in the $\l^+$ direction. This choice preserves a $U(1)\times SU(5)$
subgroup of $SO(10)$, under which a Weyl spinor $U^\a$ and an anti-Weyl
spinor $V_\a$ decompose as
\eqn\decom{U^\a \longrightarrow \(U^+_{5\over 2}, 
U_{{1\over 2}[ab]}, U^a_{-{3\over 2}}\) ,\quad
V_{\a} \longrightarrow \(V_{-{5\over 2}+}, V^{[ab]}_{-{1\over 2}}, 
V_{+{3\over 2}a}\),}
where the subscript denotes the $U(1)$ charge.

Since $(\l^+)^4$ carries $+10$ $U(1)$ charge, 
$(\lb\g^{mnp} D) A W W W$ must carry $-10$ $U(1)$ charge
which is only possible if
$(\lb\g^{mnp} D)$ carries $-3$ charge, $A_\a$ carries $-{5\over 2}$
charge, and each $W^\a$ carries $-{3\over 2}$ charge.
Contracting the $SU(5)$ indices, one finds that the unique 
$U(1)\times SU(5)$ invariant contraction of the indices is
\eqn\exone{(\l^+)^4(\lb\g_{abc} D) A_+ W^a W^b W^c.}
Returing to covariant notation, one can easily see that
\indexc\ must be proportional to the Lorentz-invariant expression
\eqn\exonea{(\lb\g_{mnp} D) (\l A) (\l \g^m W)(\l \g^n W)(\l \g^p W),}
which reduces to \exone\
in the frame where $\l^+$ is the only non-zero component of $\l^\a$.

However, to express the kinematic factor as an integral over pure spinor
superspace as in \kinet, it is convenient to have an
expression in which all $\lb_\a$'s
appear in the combination $(\l^\a \lb_\a)$.
If all $\lb$'s appear in this combination, one can use that
\eqn\pureconv{
\int d^{16} \t\int [d\l][d\lb][dr] \exp (-\l\lb - r\t)(\l\lb)^{-n} 
\l^\a \l^\b \l^\g f_{\a\b\g} }
is proportional to 
\eqn\prec{ \langle 
\l^\a \l^\b \l^\g f_{\a\b\g} \rangle.}

To convert \exonea\ to this form, it is convenient to return to
the frame in which 
$\l^+$ is the only non-zero component of $\l^\a$ and write \exone\ as
\eqn\extwo{(\l^+)^4 \e_{abcde}
(\lb^{[de]} D_+ - \lb_+ D^{[de]}) A_+ W^a W^b W^c.}
Using the superspace equations of motion for $A_\a$ and $W^\a$, it
is easy to show that 
\eqn\ides{D_+ A_+ = D_+ W^a=0, \quad
D^{[de]} A_+ + D_+ A^{[de]} =0, \quad \e_{abcde} D^{[ab]} W^c = \cF_{de}.}
So
\extwo\ is proportional to two terms which are
\eqn\exthree{(\l^+)^4 \lb_+  \e_{abcde} (D_+ A^{[de]}) W^a W^b W^c
\quad{\rm and}\quad
(\l^+)^4 \lb_+  A_+ W^a W^b  \cF_{ab}.}

The second term in \exthree\ can be easily written in covariant language
as
\eqn\exfive{(\l\lb) (\l A)(\l \g^m W)(\l \g^n W) \cF_{mn},}
which produces the desired pure spinor superspace integral of \kinet.
And the first term in \exthree\ can be written in covariant language as
\eqn\exfour{(\l\lb) \big[(\l D)(\l\g^{mn}A)\big] (\l\g^p W)(W\g_{mnp} W),}
which produces the pure spinor superspace integral
\eqn\exfive{\langle
\big[(\l D)(\l\g^{mn}A)\big] (\l\g^p W)(W\g_{mnp} W)\rangle.}

But since BRST-trivial operators decouple, 
$$\langle (\l D) 
\big[(\l\g^{mn}A)(\l\g^p W)(W\g_{mnp} W)\big]\rangle =0,$$
which implies that \exfive\ is equal to
\eqn\exsix{\langle
(\l \g^{mn} A) (\l D)\big[(\l\g^p W)(W\g_{mnp} W)\big]\rangle.}
Finally, using the superspace equation that $D_\a W^\b$
is proportional to $(\g_{mn})_\a^{\;\b} \cF^{mn}$, one finds that \exsix\
is proportional to \kinet. So the non-minimal computation
of the kinematic factor is proportional to the minimal computation of \kinet.

\subsec{Two loops}

To compute the kinematic factor at two loops using the non-minimal
prescription of \twop, first note that the regulator ${\cal N}$ can provide
22 $d_\a$ zero modes which are multiplied by the 22 zero modes of
$s^\a$. So the remaining 10 $d_\a$ zero modes must come from the
four integrated vertex operators and the three $b_\a$ ghosts.
This is only possible if each integrated vertex operators provides
a $d_\a$ zero mode through the term $(W^\a d_\a)$ and each
$b$ ghost provides two $d_\a$ zero modes through the term of 
\twod. 

After integrating over the zero modes of the dimension one fields
$(w^I_\a, \bar w^{I\a}, d^I_\a, s^{I\a})$ using the measure factors described
in \NMPS, one is left with an expression proportional
to 
\eqn\sloopone{\int d^{16}\t \int [d\l][d\lb][dr] (\l\lb)^{-6} (\l)^6 
(\lb\g^{mnp} r)^3 W W W W \exp (-\l\lb - r\t)}
\eqn\slooptwo{=\int d^{16}\t
\int [d\l][d\lb][dr] \exp (-\l\lb - r\t)(\l\lb)^{-6} (\l)^6 
(\lb\g^{mnp} D)^3 W W W W }
where the index contractions on
\eqn\indexc{(\l)^6 (\lb\g^{mnp} D)^3 W W W W }
have not been worked out.
Note that the factor of $(\l)^6$ in \sloopone\ comes from 
the $11g$ factors of $\l$ and $\lb$ which multiply the zero modes of
$d^I_\a$ and $s^I_\a$ in ${\cal N}$, the factor of $(\l)^{-8g}(\lb)^{-8g}$
in the measure factor of $w^I_\a$ and $\bar w^{I\a}$, and the factor
of $(\lb)^{-3g}$ in the measure factor of $s^{I\a}$.

As in the one-loop four-point amplitude, there is fortunately a unique
way of contracting the indices of \indexc\ in a Lorentz-invariant manner.
Choosing the Lorentz frame where $\l^+$ is the only non-zero component
of $\l^\a$, one finds that $(\l^+)^6$ contributes $+15$ $U(1)$ charge
so that each $(\lb\g^{mnp} D)$ must contribute $-3$ charge
and 
each $W$ must contribute $-{3\over 2}$ charge. Since the $-3$ component of
$(\lb \g^{mnp} D)$ is $(\lb^{[ab]} D_+ - \lb_+ D^{[ab]})$,
and since $D_+$ annihilates the $-{3\over 2}$ component of $W^\a$,
the only contribution to \indexc\ comes from a term of the form
\eqn\gexone{ (\l^+)^6 (\lb_+)^3  (D^{[ab]} D^{[cd]} D^{[ef]})
 (W^g W^h W^j W^k)}
where the ten $SU(5)$ indices are contracted with two $\e_{abcde}$'s.

The term of \gexone\ produces three types of terms depending on how
the three $D$'s act on the four $W$'s. 
If all three $D$'s act on the same $W$,
one gets a term proportional to
$(\l^+)^6 (\lb_+)^3  W W W \p \cF $, which by $U(1)\times SU(5)$ invariance
must have the form
\eqn\gexfour{ (\l^+)^6 (\lb_+)^3 W^a W^b W^c \p_a \cF_{bc}.}
And if two $D$'s act on the same $W$, one gets a term proportional to
$(\l^+)^6 (\lb_+)^3  \cF W W \p W $, which by $U(1)\times SU(5)$ invariance
must have the form
\eqn\gexthr{ (\l^+)^6 (\lb_+)^3 \cF_{bc} W^a W^b \p_a W^c.}
Finally, if each $D$
acts on a different $W$, one obtains a term that is proportional to
$(\l^+)^6 (\lb_+)^3 W \cF \cF \cF$, which by $U(1)\times SU(5)$ invariance
must have the form
\eqn\gextwo{ (\l^+)^6 (\lb_+)^3 \cF_{ab} \cF_{cd} \cF_{ef} W^f \e^{abcde}.}

The first term in \gexfour\ vanishes by Bianchi identities. And
the second term in \gexthr\ is proportional to the first term after
integrating by parts with respect to $\p_a$ and using the equation
of motion $\p_a W^a=0$. So the only contribution
to the kinematic factor comes from the third term of \gextwo, which can
be written in Lorentz-covariant notation as
\eqn\geee{
(\l\lb)^3 (\l \g^{mnpqr} \l)\cF_{mn} \cF_{pq} \cF_{rs} (\l\g^s W).}
So the non-minimal computation of the two-loop kinematic factor
agrees with the minimal computation of \kineu.

\newsec{Type-I Anomaly with Pure Spinors}

It will now be shown that the non-minimal pure spinor formalism
computation of the hexagon gauge anomaly in the Type-I superstring 
is equivalent to the RNS result of \ref\GreenQS{
	M.~B.~Green and J.~H.~Schwarz,
        ``The Hexagon Gauge Anomaly In Type I Superstring Theory,''
        Nucl.\ Phys.\ B {\bf 255}, 93 (1985).
}. As will be
shown below, the kinematic factor of the hexagon gauge variation can be
written as the pure spinor superspace integral
$$
K = \vev{(\l\g^mW^2)(\l\g^nW^3)(\l\g^pW^4)(W^5\g_{mnp}W^6)},
$$
whose bosonic part is the well-known $\e_{10}F^5$ RNS result of \GreenQS.

As discussed in
\ref\FMS{
	D.~Friedan, E.~J.~Martinec and S.~H.~Shenker,
        ``Conformal Invariance, Supersymmetry And String Theory,''
        Nucl.\ Phys.\ B {\bf 271}, 93 (1986).
} 
\ref\PolchinskiTU{
	J.~Polchinski and Y.~Cai,
        ``Consistency Of Open Superstring Theories,''
        Nucl.\ Phys.\ B {\bf 296}, 91 (1988).
}, the anomaly can be easily computed as a surface term which contributes
at the boundary of moduli space.
The result can be separated in two parts: the kinematic factor
depending only on momenta and polarizations,
and the moduli space part which depends on the worldsheet surface. 
We will be interested only in the kinematic factor, as the moduli space
part uses identical computations as in the anomaly analysis
using the RNS formalism\foot{A pedagogical presentation of these computations
can be found in \ref\liu{
	J.~Liu,
        ``Gauge And Gravitational Anomaly Cancellation In Type I SO(32) Superstring
        Theory,''
        Nucl.\ Phys.\ B {\bf 362}, 141 (1991).
}.}. 

\subsec{Kinematic factor computation}

In the type-I superstring theory with gauge group SO(N),
the massless open string six-point 
one-loop amplitude is given by
\eqn\sixpt{
{\cal A} = \sum_{top=P,NP,N}
G_{top}\int_0^{\infty} dt \langle{ {\cal N} \int dw b(w)
    (\l A_1)\prod_{r=2}^6 \int dz_r U_r (z_r)}\rangle
}
where 
$P,NP,N$
denotes the three possible
different world-sheet topologies, each of which has a different
group-factor $G_{top}$ \ref\gswII{
	Green, M. B., Schwarz, J. H., Witten, E. {\it 
	Superstring Theory: 2. Loop Amplitudes, Anomalies \& Phenomenology}.
	Cambridge University Press (1987)
}. When 
all particles are attached to one boundary, we have a cylinder with
$G_P=N\tr{(t^{a_1}t^{a_2}t^{a_3}t^{a_4}t^{a_5}t^{a_6})}$.
When particles are attached to both
boundaries, the diagram is a non-planar 
cylinder, where $G_{NP}=\tr{(t^{a_1}t^{a_2})}\tr{(t^{a_3}t^{a_4}t^{a_5}t^{a_6})}$.
And finally, there is the non-orientable M\"obius strip where
$G_{N}=-\tr{(t^{a_1}t^{a_2}t^{a_3}t^{a_4}t^{a_5}t^{a_6})}$.

We will be interested in the amplitude when all external states are massless
gluons with polarization $e^r_m$ {\it i.e.}, $a^r_m(x) = e_m^r e^{ik\cdot x}$, 
where $m=0,{\ldots} 9$ is the
space-time vector index and $r$ is the particle label
\foot{We will omit the adjoint gauge group index
from the polarizations and field-strengths for the rest of this section.}.
To probe the anomaly, 
one can compute \sixpt\ and substitute one of the external polarizations
for its respective momentum. However,
instead of first computing the six-point amplitude and substituting $e_m\rightarrow k_m$
in the answer, we will first make the gauge transformation
in \sixpt\ and then compute the 
resulting correlation function. 
This will give us the anomaly kinematic factor directly.

Under the super-Yang-Mills gauge transformation
\eqn\gauges{\d A_\a = D_\a \Omega, \quad \d A_m = \p_m \Omega,}
the integrated vertex operator $\int dz U$ changes by the surface term
$\int dz \d U = \int dz \p \Omega$, and the unintegrated vertex operator
changes by the BRST-trivial quantity $\d(\l A) = \l^\a D_\a \Omega = Q\Omega.$
Choosing $\Omega(x,\t) =e^{ik\cdot x}$ has the same effect as changing $e^m \to k^m$,
which is the desired gauge transformation to probe the anomaly.

To compute the gauge anomaly, it will be convenient to choose
the gauge transformation to act on the polarization $e_m^1$ in the
unintegrated vertex operator, so that
the gauge variation of \sixpt\ is 
\eqn\variation{
\delta {\cal A} = \sum_{top=P,N,NP}G_{top}\int_0^{\infty} dt 
                  \langle {\cal N} \int dw b(w)
                  (Q\Omega(z_1))\prod_{r=2}^6 \int dz_r U_r(z_r) \rangle.
}
``Integrating'' $Q$ by parts inside the 
correlation function will only get
a contribution from the BRST variation of the $b$-ghost,
which is
a derivative with respect to the modulus \nref\dhoker{
	E.~D'Hoker and D.~H.~Phong,
        ``The Geometry Of String Perturbation Theory,''
        Rev.\ Mod.\ Phys.\  {\bf 60}, 917 (1988).
} \refs{\FMS,\dhoker}. So
\eqn\vartwo{\eqalign{
\d{\cal A} = & -\sum_{top}G_{top} \int_0^{\infty} dt {d\over dt}
                    \langle\Omega(z_1) {\cal N} \prod_{r=2}^6
\int dz_r U_r(z_r)\rangle \cr
      \equiv & - K\sum_{top}G_{top} \Big[
	         B_{top}(\infty)-B_{top}(0)
		 \Big], 
	   }
}
where the moduli space part of the anomaly is encoded in the 
function  
$$
B_{top}(t) \equiv \int_0^{t} dz_6\int_0^{z_6} 
dz_5\int_0^{z_5} dz_4\int_0^{z_4} dz_3
                  \int_0^{z_3} dz_2 \,
                  \langle \prod_{r=1}^6 :e^{ik_r\cdot x_r}:\rangle_{top},
$$
and $K=\langle{{\cal N}U_2U_3U_4U_5U_6}\rangle$. 
From \vartwo, it is clear that the
anomaly comes from the boundary of moduli space.

To compute the kinematic factor $K$, 
observe that there is an unique way to absorb the 16 zero 
modes of $d_{\alpha}$, 11 of $s^{\a}$ and 11 of $r_{\a}$. 
The regularization factor ${\cal N}$
must provide 11 $d_{\alpha}$,
11 $s^{\alpha}$ and 11 $r_{\a}$ zero modes. The five remaining $d_{\a}$ zero modes
must come from the external vertices\foot{It follows from this
zero mode counting that the anomaly trivially vanishes for amplitudes with less
than six external massless particles.}
through $(d {\cal W})^5$. As in the computations of the previous section,
the kinematic factor is thus given by a pure spinor superspace
integral involving 3 $\l$'s
and 5 $W$'s, as can be easily verified by integrating all the zero mode measures
except $[d\l],[d\lb]$ and $[dr]$.
To find out how the indices are contracted in $K$,
choose the reference frame where only $\l^+ \neq 0$. Then one can easily check
that the unique $U(1)\times SU(5)$-invariant contraction is 
$$
K= \vev{ (\l^+)^3\e_{abcde}W_2^aW_3^bW_4^cW_5^dW_6^e
   },
$$
which in SO(10)-covariant notation translates into
\eqn\fator{
 K = \vev{ (\l\g^m W_2)(\l\g^n W_3)(\l\g^p W_4)(W_5\g_{mnp} W_6)
     }.
}

\subsec{Bosonic contribution to kinematic factor}

When all external states are gluons, there is
only one possibility to saturate the pure spinor superspace correlation
$\vev{\l^3\t^5}$. Each
superfield $W^{\a}(\t)$ must contribute one $\t$ through
the term $-{1\over 4}(\g^{mn}\t)^{\a}F_{mn}$. 
Thus, the kinematic factor \fator\ is proportional to
\eqn\kin{
\vev{ (\l\g^p\g^{m_2n_2}\t)(\l\g^q\g^{m_3n_3}\t)(\l\g^r\g^{m_4n_4}\t)
          (\t\gamma^{m_5n_5}\g_{pqr}\g^{m_6n_6}\t)}F^2_{m_2n_2}{\ldots} 
	  F^6_{m_6n_6}.
}
We will now demonstrate the equivalence with the RNS anomaly result of 
\GreenQS\ by proving that 
\eqn\imp{
\vev{(\l\g^p\g^{m_1n_1}\t)(\l\g^q\g^{m_2n_2}\t)(\l\g^r\g^{m_3n_3}\t)
     (\t\g^{m_4n_4}\g_{pqr}\g^{m_5n_5}\t)
} = {1\over 45} \e^{m_1n_1{\ldots}m_5n_5}.
}

We will first
show that the correlation in \imp\ is proportional
to $\e_{10}$ by checking its behavior
under a parity transformation.
Using the language of \PSmulti, we can rewrite \imp\ as
\eqn\compl{
(T^{-1})^{(\a\b\g)[\rho_1\rho_2\rho_3\rho_4\rho_5]}
T_{(\a\b\g)[\d_1\d_2\d_3\d_4\d_5]}(\g^{m_1n_1})^{\d_1}_{\,\,\rho_1}
(\g^{m_2n_2})^{\d_2}_{\,\,\rho_2}(\g^{m_3n_3})^{\d_3}_{\,\,\rho_3}
(\g^{m_4n_4})^{\d_4}_{\,\,\rho_4}(\g^{m_5n_5})^{\d_5}_{\,\,\rho_5},
}
where $T$ and $T^{-1}$ are defined by
\eqn\deftt{(T^{-1})^{(\a_1\a_2\a_3)[\d_1\d_2\d_3\d_4\d_5]}=
 (\g^m)^{\a_1\d_1}(\g^n)^{\a_2\d_2}(\g^p)^{\a_3\d_3}(\g_{mnp})^{\d_4\d_5}
}
$$
T_{(\a_1\a_2\a_3)[\d_1\d_2\d_3\d_4\d_5]} = 
 \g^m_{\a_1\d_1}\g^n_{\a_2\d_2}\g^p_{\a_3\d_3}(\g_{mnp})_{\d_4\d_5},
$$
and the $\a$-indices are symmetric and gamma matrix traceless, and the
$\d$-indices are antisymmetric. Since
a parity transformation has the effect of changing
a Weyl spinor $\psi^{\a}$ to an anti-Weyl spinor $\psi_{\a}$,  
it follows from the definitions of \deftt\ that a parity transformation 
exchanges $T \leftrightarrow T^{-1}$. Furthermore, since a
parity transformation also changes
$$
(\g^{mn})^{\d}_{\,\,\rho}\rightarrow (\g^{mn})_{\d}^{\,\,\rho} = 
- (\g^{mn})_{\,\,\d}^{\rho},
$$
it readily follows that the kinematic factor \compl\
is odd under parity, 
so it is proportional to $\e_{10}$. Finally, 
the proportionality constant of ${1\over {45}}$ in \imp\ can be 
explicitly computed using the identities
listed in Appendix A.

\newsec{$t_8$ and $\e_{10}$ from pure spinor superspace}

In this section, we describe some interesting identities involving
the $t_8$ and $\e_{10}$ tensors and show how they are closely related
when obtained from pure spinor superspace integrals.
This is different from computations in the RNS formalism where 
$t_8$ and $\e_{10}$ come from correlation functions with different
spin structures.

Since the one-loop $t_8 F^4$ and $\e_{10} B F_4$ 
terms are expected to be related by non-linear supersymmetry,
there might be a common superspace origin for the $t_8$ and $\e_{10}$
tensors.
This suggests looking for a BRST-closed pure spinor superspace
integral involving four super-Yang-Mills superfields whose bosonic
part involves both the $t_8$ and $\e_{10}$ tensors. One such BRST-closed 
expression we found is 
\eqn\found{\vev{(\l\g^rW^1)(\l\g^sW^2)(\l\g^tW^3)
(\t\g^m\g^n\g_{rst}W^4)}.}
Although \found\ is not spacetime supersymmetric because of the explicit
$\t$, it might be related to a supersymmetric expression
in a constant background where the $N=1$ supergravity superfield
$G_{m\a}$ satisfies $G_{m\a}=\g_{m\a\b}\t^\b + b_{mn} (\g^n\t)_\a$
for constant $b_{mn}$.

When restricted to its purely bosonic part,
\found\ 
defines the following 10-dimensional
tensor:
\eqn\fratres{
t_{10}^{mnm_1n_1m_2n_2m_3n_3m_4n_4} = \vev{ (\l\g^a \g^{m_1n_1}\t)(\l\g^b \g^{m_2n_2}\t)
	          (\l\g^c \g^{m_3n_3}\t)(\t\g^m\g^n\g_{abc}\g^{m_4n_4}\t)}.
}
Using $\g^m\g^n = \g^{mn}+\eta^{mn}$ we obtain
\eqn\idt{\eqalign{
t_{10}^{mnm_1n_1m_2n_2m_3n_3m_4n_4} =
&+ \vev{ (\l\g^a \g^{m_1n_1}\t)(\l\g^b \g^{m_2n_2}\t)
	          (\l\g^c \g^{m_3n_3}\t)(\t\g^{mn}\g_{abc}\g^{m_4n_4}\t)}\cr
& +\eta^{mn}\vev{ (\l\g^a \g^{m_1n_1}\t)(\l\g^b \g^{m_2n_2}\t)
	          (\l\g^c \g^{m_3n_3}\t)(\t\g_{abc}\g^{m_4n_4}\t)}.
}}
And using the identities listed in appendix A, one can check that\foot{The sign
in front of $\e_{10}$ depends on the chirality of $\t$. For an anti-Weyl $\t_{\a}$,
the sign is ``+''.}
\eqn\ident{
t_{10}^{mnm_1n_1m_2n_2m_3n_3m_4n_4} = 
       -  {2\over 45}\Big[ \eta^{mn}t_8^{m_1n_1m_2n_2m_3n_3m_4n_4}
                      - {1\over 2}\e^{mnm_1n_1m_2n_2m_3n_3m_4n_4}\Big]
}
where the $t_8$ tensor is defined as usual by its contraction
with 
four field-strengths to give
$$
\eqalign{t_8^{m_1n_1{\ldots} m_4n_4}F^1_{m_1n_1}{\ldots} F^4_{m_4n_4} =
  & + 8(F^1F^2F^3F^4) + 8(F^1F^3F^2F^4) + 8(F^1F^3F^4F^2) \cr
  & -2(F^1F^2)(F^3F^4) - 2(F^2F^3)(F^4F^1) - 2(F^1F^3)(F^2F^4).
}  
$$

It is also interesting to contrast the similarity between $\e_{10}$ and
$t_8$ when written in terms of the $T$ and $T^{-1}$ tensors:
\eqn\simi{\eqalign{
\epsilon^{mnm_1n_1{\ldots}m_4n_4} & \propto
(T^{-1})^{(\a\b\g)[\rho_0\rho_1\rho_2\rho_3\rho_4]}
T_{(\a\b\g)[\d_0\d_1\d_2\d_3\d_4]}(\g^{mn})^{\d_0}_{\,\,\rho_0}
{\ldots} 
(\g^{m_4n_4})^{\d_4}_{\,\,\rho_4} \cr
t_8^{m_1n_1{\ldots}m_4n_4} & \propto
(T^{-1})^{(\a\b\g)[\kappa\rho_1\rho_2\rho_3\rho_4]}
T_{(\a\b\g)[\kappa\d_1\d_2\d_3\d_4]}
(\g^{m_1n_1})^{\d_1}_{\,\,\rho_1}{\ldots} 
(\g^{m_4n_4})^{\d_4}_{\,\,\rho_4},
}}
which shows, in a pure spinor superspace language, how one can ``obtain''
the $t_8$ tensor from $\e_{10}$: it is a matter of removing 
$(\g^{mn})^{\d_0}_{\,\,\rho_0}$ and contracting the associated spinorial
indices in $T$ and $T^{-1}$. 
So when using pure spinors, there is a close relation between these
two different-looking tensors.

\vskip 15pt
{\bf Acknowledgements:} 
CRM acknowledges FAPESP grant 04/13290-8 for financial support and
NB acknowledges  CNPq grant 300256/94-9, Pronex
grant 66.2002/1998-9, and FAPESP grant 04/11426-0 for partial financial
support.

\appendix{A}{Pure Spinor Superspace Identities}

In this appendix we list all the identities used throughout this
paper. They were obtained with 
the inestimable help of Ulf Gran's GAMMA package \ref\GAMMA{
	U.~Gran,
        ``GAMMA: A Mathematica package for performing Gamma-matrix algebra and  Fierz
        transformations in arbitrary dimensions,''
        arXiv:hep-th/0105086.
}
along with some custom-made functions to handle $\e_{10}$ tensors.
The convention used for antisymmetrization of $n$ indices 
is that one must divide by $n!$. Furthermore, it is sometimes 
more convenient to use the notation $\d^{a_1{\ldots} a_n}_{m_1{\ldots} m_n}=
\d^{[a_1}_{m_1}{\ldots}\d^{a_n]}_{m_1}$, e.g.,
$$
\d^{a_1a_2}_{m_1m_2} = {1\over 2!}\( \d^{a_1}_{m_1}\d^{a_2}_{m_2}
 - \d^{a_2}_{m_1}\d^{a_1}_{m_2} \),
$$
and -- for notational simplicity -- not care about the difference 
between downstairs and upstairs indices in the formul{\ae}.

\subsec{Identities ad nauseam}

The computation of a correlation like 
$$
\vev{ (\l\g^m\g^{m_1n_1}\t)(\l\g^n\g^{m_2n_2}\t)(\l\g^p\g^{m_3n_3}\t)
(\t\g^{m_4n_4}\g_{mnp}\g^{m_5n_5}\t)}
$$
or
$$
\vev{(\l\g^{m}\t)(\l\g^a \g^{m_1n_1}\t)(\l\g^{bcn} \g^{m_2n_2}\t)
(\t\g^{m_3n_3}\g_{abc}\g^{m_4n_4}\t)}
$$
requires a lot of identities, which will be listed below.

We first define $(\t\gamma^{m_4n_4}\g_{mnp}\g^{m_5n_5}\t)=
G^{m_4n_4m_5n_5}_{mnpr_1r_2r_3}(\t\g^{r_1r_2r_3}\t)$. One can check
that
\eqn\gtensor{
G^{m_4n_4m_5n_5}_{mnpr_1r_2r_3}=
	+ {1\over 6}\epsilon^{mm_4m_5nn_4n_5pr_1r_2r_3}
        - 24\d^{np}_{n_4n_5}\d^{mm_4m_5}_{r_1r_2r_3} 
	+ 12\d^{m_5n_5}_{n_4p}\d^{mm_4n}_{r_1r_2r_3} 	
}
$$
- 6\d^{m_5n_5}_{np}\d^{mm_4n_4}_{r_1r_2r_3} +
  12\d^{m_4n_4}_{n_5p}\d^{mm_5n}_{r_1r_2r_3} 
- 6\d^{m_4n_4}_{np}\d^{mm_5n_5}_{r_1r_2r_3} 
 - 2\d^{m_4n_4}_{m_5n_5}\d^{mnp}_{r_1r_2r_3} + [mnp] +[m_4n_4] + [m_5n_5],
$$
and $+[mnp]+[m_4n_4]+[m_5n_5]$ means that one must antisymmetrize in those indices.

The computation $t_8$ also requires the 
identity $(\t\g^{abc}\g^{mn}\t)=(\t\g^{r_1r_2r_3}\t)K^{abcmn}_{r_1r_2r_3}$,
where
$$
K^{abcmn}_{r_1r_2r_3} = 
-\eta^{cn}\d^{abm}_{r_1r_2r_3} + 
 \eta^{cm}\d^{abn}_{r_1r_2r_3} + 
 \eta^{bn}\d^{acm}_{r_1r_2r_3} 
-\eta^{bm}\d^{acn}_{r_1r_2r_3} - 
 \eta^{an}\d^{bcm}_{r_1r_2r_3} + 
 \eta^{am}\d^{bcn}_{r_1r_2r_3}
$$
The following identity is also useful\foot{This identity was suggested by
Pierre Vanhove during discussions of \PSequivI.}
\eqn\defi{\eqalign{
          (\l\g^{mnp}\t)(\l\g^{qrs}\t) 
          & = -{1\over 96}(\t\g^{tuv}\t)(\l\g^{mnp}\g_{tuv}\g^{qrs}\l) \cr
          & \equiv - {1\over 96}(\l\g^{abcde}\l)(\t\g^{tuv}\t)f^{mnpqrs}_{abcdetuv}
	  }
}
where
\eqn\ftensor{
f^{mnpqrs}_{abcdetuv} = 
18(\d^{rs}_{uv}\d^{abcde}_{mnpqt} - \d^{np}_{uv}\d^{abcde}_{qrsmt}
 + \d^{mn}_{qr}\d^{abcde}_{pstuv})
}
$$
+54(\d^{nv}_{rs}\d^{abcde}_{mpqtu}-\d^{rv}_{np}\d^{abcde}_{qsmtu}
+\d^{ps}_{tv}\d^{abcde}_{mnqru})+[mnp]+[qrs]+[tuv].
$$

Using the gamma matrix identities
$$
(\l \g^m\g^{np}\t) = (\l\g^{mnp}\t) +\eta^{mn}(\l\g^p\t)
-\eta^{mp}(\l\g^n\t),
$$
$$
(\l\g^{abc}\g^{de}\t) = + (\l\g^{abcde}\t)
-2\d^{bc}_{de}(\l\g^{a}\t) 
+ 2\d^{ac}_{de}(\l\g^{b}\t) 
- 2\d^{ab}_{de}(\l\g^{c}\t)
$$
$$
- \d^{c}_{e}(\l\g^{abd}\t) + 
 \d^{c}_{d}(\l\g^{abe}\t) + 
 \d^{b}_{e}(\l\g^{acd}\t) - 
 \d^{b}_{d}(\l\g^{ace}\t) - 
 \d^{a}_{e}(\l\g^{bcd}\t) + 
 \d^{a}_{d}(\l\g^{bce}\t) 
$$
and the definitions above, all correlations considered in this paper
turn into a
linear combination of the following building-blocks:
\eqn\unus{
\langle (\l\g^m\t)(\l\g^n\t)(\l\g^p\t)(\t\g_{ijk}\t)\rangle = {1\over 120}
\delta^{mnp}_{ijk}
}
\eqn\duo{
\langle(\l\g^{mnp}\t)(\l\g_{q}\t)(\l\g_{t}\t)(\t\g_{ijk}\t)\rangle=
       {1\over 70}\d^{[m}_{[q}\eta_{t][i}\d^{n}_j\d^{p]}_{k]}
}
\eqn\tres{
\langle(\l\g_{t}\t)(\l\g^{mnp}\t)(\l\g^{qrs}\t)(\t\g_{ijk}\t)\rangle=
{1\over 8400}\e^{ijkmnpqrst}+
}
$$
+{1\over 140}\Big[ 
	 \d^{[m}_t\d^n_{[i}\eta^{p][q}\d^r_j\d^{s]}_{k]}
	-\d^{[q}_t\d^r_{[i}\eta^{s][m}\d^n_j\d^{p]}_{k]}
\Big]
-{1\over 280}\Big[ 
	 \eta_{t[i}\eta^{v[q} \d^r_j\eta^{s][m}\d^n_{k]}\d^{p]}_v 
	-\eta_{t[i}\eta^{v[m} \d^n_j\eta^{p][q}\d^r_{k]}\d^{s]}_v
\Big].
$$
\eqn\quattuor{
\langle (\l\g^{mnpqr}\t)(\l\g_{stu}\t)(\l\g^v\t)(\t\g_{fgh}\t)=
{1\over 35}
\eta^{v[m}\d^n_{[s}\d^p_t \eta_{u][f}\d^q_g\d^{r]}_{h]}
-{2\over 35}
\d^{[m}_{[s}\d^n_t\d^p_{u]}\d^q_{[f}\d^{r]}_g\d^v_{h]}
}
$$
+{1\over 120}\e^{mnpqr}_{\qquad\;\; abcde}\left(
{1\over 35}
\eta^{v[a}\d^b_{[s}\d^c_t \eta_{u][f}\d^d_g\d^{e]}_{h]}
-{2\over 35}
\d^{[a}_{[s}\d^b_t\d^c_{u]}\d^d_{[f}\d^{e]}_g\d^v_{h]}
\right)
$$
\eqn\ind{\langle(\l\g^{mnpqr}\l)(\l\g^{u}\t)(\t\g_{fgh}\t)
(\t\g_{jkl}\t)\rangle = 
}
$$
-{4\over 35}{\Big[}\d^{[m}_{[j} \d^n_k \d^p_{l]}\d^q_{[f}\d^{r]}_g \d^u_{h]}
+\d^{[m}_{[f} \d^n_g \d^p_{h]}\d^q_{[j}\d^{r]}_k \d^u_{l]}
-{1\over 2}\d^{[m}_{[j}\d^n_k \eta_{l][f}\d^p_g\d^q_{h]}\eta^{r]u}
-{1\over 2}\d^{[m}_{[f}\d^n_g \eta_{h][j}\d^p_k\d^q_{l]}\eta^{r]u}
\Big]
$$
$$
-{1\over 1050}\e^{mnpqr}_{\qquad\;\; abcde}{\Big[}
\d^{[a}_{[j} \d^b_k \d^c_{l]}\d^d_{[f}\d^{e]}_g \d^u_{h]}
+\d^{[a}_{[f} \d^b_g \d^c_{h]}\d^d_{[j}\d^{e]}_k \d^u_{l]}
$$
$$
-{1\over 2}\d^{[a}_{[j}\d^b_k \eta_{l][f}\d^c_g\d^d_{h]}\eta^{e]u}
-{1\over 2}\d^{[a}_{[f}\d^b_g \eta_{h][j}\d^c_k\d^d_{l]}\eta^{e]u}
\Big]
$$
\eqn\quinque{
\langle (\l\g^{mnpqr}\t)(\l\g_{d}\t)(\l\g_e\t)(\t\g_{fgh}\t)=
 - {1\over 42}\delta^{mnpqr}_{defgh} 
 - {1\over 5040}\e^{mnpqr}_{\qquad\;\; defgh}
}
\eqn\tind{
\langle(\l\g^{mnpqr}\l)(\l\g^{stu}\t)(\t\g_{fgh}\t)
(\t\g_{jkl}\t)\rangle=}
$$
-{12\over 35}\Big[
	\d^{[s}_{[f}\d^t_g \eta^{u][m}\d^n_{h]}\d^p_{[j}
	\d^q_k\d^{r]}_{l]}
      + \d^{[s}_{[j}\d^t_k \eta^{u][m}\d^n_{l]}\d^p_{[f}
        \d^q_g\d^{r]}_{h]} 
$$
$$
      - \eta^{v[s}\d^t_{[f}\eta^{u][m}\d_g^n\eta_{h][j}
        \d^p_k\d^q_{l]}\d_v^{r]}
      - \eta^{v[s}\d^t_{[j}\eta^{u][m}\d_k^n\eta_{l][f}
        \d^p_g\d^q_{h]}\d_v^{r]} 
\Big]
$$
$$
-{1\over 350}\e^{mnpqr}_{\qquad \, abcde}\Big[
	\d^{[s}_{[f}\d^t_g \eta^{u][a}\d^b_{h]}\d^c_{[j}\d^d_k\d^{e]}_{l]}
      + \d^{[s}_{[j}\d^t_k \eta^{u][a}\d^b_{l]}\d^c_{[f}\d^d_g\d^{e]}_{h]} 
$$
$$
      - \eta^{v[s}\d^t_{[f}\eta^{u][a}\d_g^b\eta_{h][j}\d^c_k
        \d^d_{l]}\d_v^{e]}
      - \eta^{v[s}\d^t_{[j}\eta^{u][a}\d_k^b\eta_{l][f}\d^c_g
        \d^d_{h]}\d_v^{e]} 
\Big].
$$
\eqn\septem{
\vev{(\l\g^{mnp}\t)(\l\g^{qrs}\t)(\l\g_{tuv}\t)(\t\g_{ijk}\t)}=
}
$$
-{3\over 175}\Big[- \d^{[i}_a\d^{j}_{[q}\d^{k]}_{r}\d^{[m}_{s]}\d^{n}_{[t}\d^{p]}_{u}\d^a_{v]}
+ \d^{[i}_a\d^{j}_{[t}\d^{k]}_{u}\d^{[m}_{v]}\d^{n}_{[q}\d^{p]}_{r}\d^a_{s]} 
+ \d^{[i}_{[q}\d^{j}_{r}\eta^{k][m}\eta_{s][t}\d^{n}_{u}\d^{p]}_{v]}
$$
$$
+\d^a_{[t}\eta^{b[i}\d^j_u\eta^{k][m}\eta_{v][q}\d^n_r\eta_{s]a}\d^{p]}_b
- \d^{a}_{[q}\eta^{b[i}\d^{j}_{r}\eta^{k][m}\eta_{s][t}\d^{n}_{u}\eta_{v]a}\d^{p]}_b 
- \d^{[i}_{[t}\d^{j}_{u}\eta^{k][m}\eta_{v][q}\d^{n}_{r}\d^{p]}_{s]}
\Big]
$$
$$
+{1\over 33600}\e^{abcde}_{\qquad \, a_1a_2a_3a_4a_5}
f^{mnpqrs}_{abcdefgh}\Big[
	\d^{[t}_{[f}\d^u_g \eta^{v][a_1}\d^{a_2}_{h]}\d^{a_3}_{[i}\d^{a_4}_j\d^{a_5]}_{k]}
      + \d^{[t}_{[i}\d^u_j \eta^{v][a_1}\d^{a_2}_{k]}\d^{a_3}_{[f}\d^{a_4}_g\d^{a_5]}_{h]} 
$$
$$
      - \eta^{z[t}\d^u_{[f}\eta^{v][a_1}\d_g^{a_2}\eta_{h][i}\d^{a_3}_j
        \d^{a_4}_{k]}\d_z^{a_5]}
      - \eta^{z[t}\d^u_{[i}\eta^{v][a_1}\d_j^{a_2}\eta_{k][f}\d^{a_3}_g
        \d^{a_4}_{h]}\d_z^{a_5]} 
\Big].
$$
These identities
can be straightforwardly derived. The recipe is the following.
One writes the most general 
tensor containing Kronecker deltas with the same
symmetry properties as the left hand side  and then
contracts some appropriate indices to find the coefficients 
which satisfy the normalization
$\vev{\l^3\t^5}=1$. After obtaining all terms containing only
Kronecker deltas one can find terms with 
$\e_{10}$ tensors considering the duality properties of the gamma matrices:
$$
\(\g^{m_1m_2m_3m_4m_5}\)_{\a\b}=+{1\over 5!}
\e^{m_1m_2m_3m_4m_5n_1n_2n_3n_4n_5}\(\g_{n_1n_2n_3n_4n_5}\)_{\a\b},
$$
$$
\(\g^{m_1m_2m_3m_4m_5m_6}\)^{\,\,\b}_{\a}=+{1\over 4!}
\e^{m_1m_2m_3m_4m_5m_6n_1n_2n_3n_4}\(\g_{n_1n_2n_3n_4}\)^{\,\,\b}_{\a},
$$
$$
\(\g^{m_1m_2m_3m_4m_5m_6m_7}\)_{\a\b}=-{1\over 3!}
\e^{m_1m_2m_3m_4m_5m_6m_7n_1n_2n_3}\(\g_{n_1n_2n_3}\)_{\a\b},
$$
$$
\(\g^{m_1m_2m_3m_4m_5m_6m_7m_8}\)^{\,\,\b}_{\a}=-{1\over 2!}
\e^{m_1m_2m_3m_4m_5m_6m_7m_8n_1n_2}\(\g_{n_1n_2}\)^{\,\,\b}_{\a}.
$$

The following identities turn out to be useful when doing all these 
manipulations and can be derived using the properties of pure spinors and
gamma matrices:
\eqn\gru{
\(\g^{mnp}\)_{\a\b}\(\g_{mnp}\)^{\g\d} = 48\( \d^{\g}_{\a}\d^{\d}_{\b} -
      \d^{\g}_{\b}\d^{\d}_{\a}\), \hskip 0.2in
(\l\g_m \psi)(\l\g^m \xi)=0 \quad \forall \psi^{\a},\; \xi^{\a}
}
\eqn\grd{
  (\l\g^{mnpqr}\l)(\l\g_{mna}\t) = 0, \hskip 0.2in
  (\l\g^{mnpqr}\l)(\l\g_{m}\t)=0
}
\eqn\grt{
  (\l\g^{amn}\t)(\l\g_a\t)= 2(\l\g^m\t)(\l\g^n\t), \hskip 0.2in
  (\l\g^{abm}\t)(\l\g^{abn}\t)= -4(\l\g^m\t)(\l\g^n\t)
}
\eqn\grq{
  (\l\g^{mabcn}\l)(\t\g_{abc}\t)=96(\l\g^m\t)(\l\g^n\t),
}
\eqn\grqq{
  (\l\g^{abcmn}\t)(\l\g_{abc}\t)=-36(\l\g^m\t)(\l\g^n\t),
}
\eqn\grc{
  (\l\g^a\g^{bcmn}\t)(\l\g_{abc}\t)=-28(\l\g^m\t)(\l\g^n\t),
}
\eqn\grs{\eqalign{
(\l\g^{abc}\t)(\l\g^{ade}\t) =& -(\l\g^{cde}\t)(\l\g^b\t) + (\l\g^{bde}\t)(\l\g^c\t)
                                + (\l\g^{bce}\t)(\l\g^d\t) \cr
                             & - (\l\g^{bcd}\t)(\l\g^e\t) - \eta^{ce}(\l\g^b\t)(\l\g^d\t) 
                               + \eta^{cd}(\l\g^b\t)(\l\g^e\t) \cr
                             & + \eta^{be}(\l\g^c\t)(\l\g^d\t)-\eta^{bd}(\l\g^c\t)(\l\g^e\t)
}}
$$
\eqalign{
  (\l\g^{abcde}\t)(\l\g^{agh}\t) = & +(\l\g^{hbcde}\t)(\l\g^{g}\t) 
        - (\l\g^{gbcde}\t)(\l\g^h\t) + (\l\g^{bcd}\t)(\l\g^{egh}\t) \cr
	& - (\l\g^{bce}\t)(\l\g^{dgh}\t) + (\l\g^{bde}\t)(\l\g^{cgh}\t) 
	    - (\l\g^{cde}\t)(\l\g^{bgh}\t) \cr
	& -\eta^{he}(\l\g^{bcd}\t)(\l\g^g\t) + \eta^{hd}(\l\g^{bce}\t)(\l\g^g\t)
	    - \eta^{hc}(\l\g^{bde}\t)(\l\g^g\t) \cr
	& + \eta^{hb}(\l\g^{cde}\t)(\l\g^g\t) + \eta^{ge}(\l\g^{bcd}\t)(\l\g^h\t) 
	    - \eta^{gd}(\l\g^{bce}\t)(\l\g^h\t) \cr
	& + \eta^{gc}(\l\g^{bde}\t)(\l\g^h\t) - \eta^{gb}(\l\g^{cde}\t)(\l\g^h\t)
}
$$
$$
\eqalign{
(\l\g^{abcde}\t)(\l\g^{abh}\t) = & -4 \eta^{eh}(\l\g^c\t)(\l\g^d\t) 
   + 4 \eta^{dh}(\l\g^c\t)(\l\g^e\t) - 4 \eta^{hc}(\l\g^d\t)(\l\g^e\t) \cr
   & - 2 (\l\g^{cde}\t)(\l\g^h\t)
}
$$

\subsec{Other pure spinor representations for $t_8$ and $\e_{10}$}

The following correlations also give rise 
to identities for $t_8$ and $\e_{10}$,
$$
\vev{(\l\g^{m}\t)(\l\g_a W^1)(\l\g_b W^2)(W^3\g^{abn}W^4)}
+{\rm perm}(1234),
$$
$$
\vev{(\l\g^a W^1)(\l\g^b W^2)(\l\g^n W^3)
    (\t\g^m\g_{ab}W^4)}+{\rm perm}(1234).
$$
Indeed one can show that
$$
\vev{(\l\g^{[m|}\t)(\l\g_a \g^{m_1n_1}\t)(\l\g_b \g^{m_2n_2}\t)
(\t\g^{m_3n_3}\g^{ab|n]}\g^{m_4n_4}\t)} 
+{\rm p}(1234) = -{116\over 525}\e^{mnm_1n_1{\ldots}m_4n_4}
$$
$$
\eta_{mn}\vev{(\l\g^m\t)(\l\g_a \g^{m_1n_1}\t)(\l\g_b \g^{m_2n_2}\t)
(\t\g^{m_3n_3}\g^{abn}\g^{m_4n_4}\t)} 
+{\rm p}(1234) = {16\over 15}t_8^{m_1n_1{\ldots}m_4n_4}
$$
$$
\vev{(\l\g^a \g^{m_1n_1}\t)(\l\g^b \g^{m_2n_2}\t)(\l\g^{[n} \g^{m_3n_3}\t)
    (\t\g^{m]}\g_{ab}\g^{m_4n_4}\t)}
    + {\rm p}(1234) = {2\over 175}\e^{mnm_1n_1{\ldots}m_4n_4}
$$
$$
\eta_{mn}\vev{(\l\g^a \g^{m_1n_1}\t)(\l\g^b \g^{m_2n_2}\t)(\l\g^{n} \g^{m_3n_3}\t)
    (\t\g^{m}\g_{ab}\g^{m_4n_4}\t)}+{\rm p}(1234) = -{16\over 15}t_8^{m_1n_1{\ldots}m_4n_4}.
$$

\listrefs

\end